\documentclass[12pt]{article}
\usepackage[margin=1in]{geometry}
\usepackage{authblk}
\usepackage{graphicx, subcaption, booktabs}
\usepackage{placeins, xspace, hyperref}
\usepackage{amsmath, amssymb, siunitx}
\usepackage[version=4]{mhchem}

\newcommand{\Y}{\ce{_{39}^{97}Y}\xspace}
\newcommand{\I}{\ce{_{53}^{136}I}\xspace}

\title{Xe gas bubble re-solution in U-10Mo nuclear fuel}

\author[1]{ATM Jahid Hasan}
\author[2]{Linu Malakkal}
\author[2]{Mathew Swisher}
\author[1,2,\thanks{Corresponding author. Email: bwbeeler@ncsu.edu.}]
{Benjamin~Beeler}
\affil[1]{North Carolina State University}
\affil[2]{Idaho National Laboratory}

\begin{document}

\maketitle

\begin{abstract}
	The U.S. High-Performance Research Reactor program
	aims to convert high-power research reactors
	from highly enriched uranium to low-enriched uranium
	using a monolithic U-10Mo fuel design.
	A critical aspect of U-10Mo fuel performance is fission gas bubble behavior.
	These bubbles grow by trapping gas atoms (particularly Xe)
	but can disintegrate via irradiation-induced ``re-solution''.
	The interplay between the trapping and re-solution rates
	governs bubble evolution, impacting fuel performance and safety.
	In this study,
	binary collision approximation (BCA) and molecular dynamics (MD) simulations
	were performed to quantify the Xe gas bubble re-solution rate in U-10Mo fuel.
	First, the energy loss of fission fragments (FFs)
	through electronic and nuclear stopping was evaluated.
	The effect of electronic stopping on re-solution was then analyzed
	using MD simulations coupled with the two-temperature model.
	Results indicate that thermal spikes generated by electronic stopping
	do not contribute to gas bubble re-solution in U-10Mo.
	To quantify re-solution due to nuclear stopping,
	BCA simulations of FFs in U-10Mo were performed
	to obtain the average FF incidence probability, energy, and angle
	as a function of distance from the FF origin.
	Subsequent simulations assessed FF--bubble interactions in U-10Mo
	for different FF energies and bubble radii.
	From these analyses, an overall re-solution rate $b$ was calculated
	at equilibrium bubble pressure per unit fission rate density,
	yielding values ranging
	from \num{4.4e-26} m$^3$/fission for the largest bubbles
	to \num{8.8e-25} m$^3$/fission for the smallest.
	The effect of bubble pressure on the re-solution rate was also evaluated,
	revealing an inverse relationship between the two.
\end{abstract}

\section{Introduction}

A U-10Mo alloy-based monolithic fuel design was selected
as the fuel type for converting U.S. High-Performance Research Reactors (HPRRs)
\cite{meyer2014} from highly enriched uranium (HEU) fuel
to low-enriched (LEU) fuel.
To reliably predict the fuel's behavior under irradiation,
mesoscale and engineering-level fuel performance models require
knowledge of the fundamental mechanistic behavior of fission products
within the fuel to describe key phenomena,
such as swelling \cite{beeler2018gb, annualreport2021}.
Specifically, understanding the progression of Xe gas bubbles in the fuel
is crucial for optimizing reactor performance and safety.
These gas bubbles act as a sink for diffusing Xe atoms in the fuel,
whose subsequent entrapment leads to progressive bubble growth.
Conversely, under irradiation, the Xe atoms in the gas bubbles are reintroduced
into the fuel matrix through collision cascades and thermal spikes
produced by fission fragments (FFs)---a process referred to as ``re-solution''.
The relative rates of Xe trapping and re-solution
dictate the size and density of the gas bubbles
\cite{ye2023, olander2006re, parfitt2008},
thereby influencing bubble evolution
and the overall swelling behavior of the fuel.

There are two widely accepted mechanisms of fission gas bubble re-solution:
homogeneous re-solution and heterogeneous re-solution \cite{olander2006re}.
In the homogeneous model proposed by Nelson \cite{nelson1969},
individual atoms are ejected from gas bubbles via collisions
with energetic FFs or recoil atoms traversing the bubbles.
These atomic collision cascades are primarily
governed by the nuclear stopping power of the material.
In contrast,
in the heterogeneous model proposed by Turnbull \cite{turnbull1971},
a portion of the gas bubbles is dissolved by FFs passing in their vicinity.
The driving mechanism is
the local heating of the material containing the gas bubbles,
through the electronic stopping of the FFs \cite{setyawan2018}.
Irrespective of the mechanism,
the re-solution rate $b$ is defined as
the fraction of gas atoms returned to the solid solution
from bubbles per unit time,
or equivalently, the probability per unit time of
a single gas atom being ejected from a bubble back into the lattice
\cite{olander2006re, setyawan2018}.
Since both mechanisms occur on very short timescales,
experimentally determining the re-solution rates
required for fission gas release models becomes challenging.
As a result, atomic-scale simulations are essential
to elucidate the underlying mechanisms
and provide a quantitative description of the re-solution process in U-10Mo.

In the literature up to this point,
atomistic simulations have been widely used
to evaluate the re-solution rate in various nuclear materials.
For instance, in 2008, Parfitt et al. \cite{parfitt2008}
simulated primary knock-on atoms (PKAs) in uranium dioxide (UO$_2$)
using molecular dynamics (MD) to assess the re-solution of helium gas bubbles.
In 2009, Schwen et al. \cite{schwen2009md} investigated
the homogeneous re-solution of Xe gas bubbles in UO$_2$,
using binary collision approximation (BCA) and MD.
The following year, Huang et al. \cite{huang2010md}
examined the impact of thermal spikes on Xe re-solution in UO$_2$
using MD coupled with the two-temperature model (TTM).
In 2012, Govers et al. \cite{govers2012} performed MD simulations to study
how PKA and thermal spikes interact with Xe gas bubbles in UO$_2$
and proposed a mathematical model to describe the observed re-solution.
The most comprehensive work on Xe gas bubble re-solution in UO$_2$
was conducted in 2018 by Setyawan et al. \cite{setyawan2018}.
They reconciled the inconsistencies found in the conclusions of previous works
on Xe bubble re-solution in UO$_2$
and evaluated the re-solution rate as a function of bubble radius
with the help of extensive MD simulations.
Their results suggest that heterogeneous re-solution of gas bubbles
is the dominant method of re-solution in UO$_2$.
In addition to UO$_2$,
the re-solution rate of fission gas bubbles has also been evaluated
in other nuclear materials.
Matthews et al. \cite{matthews2015} evaluated re-solution
in uranium carbide (UC) in 2015,
while Mao et al. \cite{mao2025} studied re-solution
in uranium-zirconium (U-Zr) alloys in 2025,
both using BCA simulations.
Unlike with UO$_2$,
thermal spikes are not expected to occur in UC and U-Zr systems,
due to their higher electronic conductivities and thermal diffusivities
\cite{ronchi1986, matthews2015, mao2025}.
Thus, only homogeneous re-solution has been studied in UC and U-Zr.
In summary, both BCA and MD simulations have been utilized
to determine re-solution rates in nuclear fuels.

MD simulations of homogeneous re-solution typically involve
assigning a high kinetic energy to a regular lattice atom
in order to emulate a PKA.
The PKA then interacts ballistically with other atoms,
initiating a collision cascade near the gas bubble
and inducing localized atomic disorder \cite{parfitt2008, govers2012}.
One alternative MD approach focuses on simulating a subcascade
by imparting energy directly to a random gas atom within the bubble.
This method reduces computational costs
by avoiding unnecessary cascade events
that may not significantly influence the re-solution process.
However, to implement this approach accurately,
BCA simulations are required in order to first obtain
an energy spectrum of gas atom PKAs \cite{schwen2009md}.
One challenge in modeling homogeneous re-solution with MD
is the channeling of PKAs or their recoils over long distances,
without undergoing significant collisions \cite{jarrin2021}.
This phenomenon can make it computationally demanding
to gather statistics on interactions between PKAs and gas bubble atoms,
particularly when PKA directions are assigned randomly.
A potential solution is to direct the PKAs
along high-index lattice directions \cite{stoller2000},
thus increasing the collision probability.
Additionally,
collision cascades often produce heat spikes due to nuclear stopping,
which induce defect formation and facilitate damage-assisted re-solution.
Ballistically re-solved atoms can then be differentiated
by employing a threshold atomic speed
that is highly improbable to occur in normal thermal equilibrium of the bubble
at the lattice temperature prior to the cascade initiation
\cite{parfitt2008}.

For heterogeneous re-solution,
MD simulations of thermal spikes are commonly employed.
FFs lose a significant portion of their energy via electronic stopping,
with the deposited energy initially raising the temperature
of the electronic subsystem.
The energy is subsequently transferred to the lattice as thermal energy
via electron-phonon coupling,
leading to a rapid rise of lattice temperature within a cylindrical zone
of typically a few nanometers in radius.
This localized heating, known as a thermal spike
\cite{wang1994, toulemonde2002, patra2019},
can induce re-solution if the spike intersects a gas bubble.
Although MD simulations cannot model electronic interactions directly,
the thermal spike process can be approximated by either
instantaneously increasing the temperature of atoms in a cylindrical region
\cite{govers2012, setyawan2018}
or by coupling MD with the TTM \cite{duffy2006, huang2010md}.

Accurate prediction of fission gas behavior
under various operational and transient conditions is critical
for the qualification of U-10Mo fuel.
To this end, the Dispersion Analysis Research Tool (DART),
a mesoscale code developed by Argonne National Laboratory \cite{ye2023},
has been equipped with the ability to calculate
fission gas swelling in U-10Mo under a range of operating conditions.
DART employs a re-solution model
that includes a piecewise function to account for
both intragranular (typically with radii $\le 2.5$ nm)
and intergranular gas bubbles (typically with radii $> 2.5$ nm) \cite{kim2008}.
However, the parameters of this model are calibrated
by fitting the computed swelling values to experimental data,
providing only a rough estimation of the re-solution rate.
A physics-based model of the re-solution rate would enhance
the predictive capability of the higher-length-scale swelling models.
In this study, we utilized BCA and MD simulations
to investigate the re-solution of Xe gas bubbles in U-10Mo fuel,
addressing both the homogeneous and heterogeneous re-solution mechanisms.
This work intends to provide
a mechanistic understanding of the re-solution process in U-10Mo,
enabling more rigorous modeling of fission gas behavior and swelling
in the fuel under various reactor conditions.

\section{Computational methods}

In this section, we outline general computational methods used in this study.
For better readability,
specific simulation parameters and implementation details
are discussed along with the results.

\subsection{Binary collision approximation}

RustBCA, an open-source software for simulating ion-material interactions
\cite{drobny2021},
was used for all BCA simulations in this work.
RustBCA supports a variety of ion-material interactions,
including sputtering, implantation, and reflection.
Out of the box, RustBCA supports infinite homogeneous 0D targets,
finite-depth layered inhomogeneous 1D targets,
inhomogeneous 2D targets through a triangular mesh,
and homogeneous 3D triangular mesh geometry.
However, it does not support inhomogeneous 3D geometry,
which is necessary to emulate gas bubbles embedded in solids.
For this reason, we implemented a custom 3D geometry in RustBCA
termed \texttt{SPHEREINCUBOID}.
This configuration enables the user to specify
a spherical material inside a distinct cuboid material.
The source code for this implementation is available
at \url{https://github.com/ATM-Jahid/RustBCA}.
Like most BCA codes, RustBCA assumes an amorphous, static material,
neglecting crystal structures and accumulation of irradiation damage.
It also cannot take into account temperature effects.

Electronic stopping in the BCA simulations performed in this work
was described by the Biersack-Varelas interpolation \cite{varelas1970},
and the nuclear interactions were described by the universal Kr-C potential
\cite{moller1984, eckstein2013}.
An exponentially distributed mean-free-path model was used for gaseous regions
and a constant mean-free-path model for solid regions.
A threshold number density of $\num{1.5e28}$ m$^{-3}$ was used to distinguish
between the gaseous and solid regions.

\subsection{Molecular dynamics}

The LAMMPS software package \cite{lammps} was utilized
to perform MD simulations,
using a U-Mo-Xe angular-dependent potential
\cite{smirnova2013, starikov2018, beelerADP}.
This angular-dependent potential can accurately describe
the body-centered cubic (bcc) phase of $\gamma$U-Mo alloys,
effectively reproducing their stable structure, elastic modulus,
room temperature density, and melting point.
To emulate the electronic stopping of the FFs,
all MD simulations were coupled with the TTM
using the \texttt{ttm/mod} command in LAMMPS \cite{norman2013, pisarev2014}.
This approach treats the electronic subsystem as a continuum
while describing the ionic subsystem through standard MD.
Energy transfer within the electronic subsystem
is governed by the heat diffusion equation,
which includes source terms to model the heat transfer
between the electronic and ionic subsystems:
\begin{align}
	C_e (T_e) \rho_e \frac{\partial T_e}{\partial t}
		&= \nabla (\kappa_e \nabla T_e) + g_p (T_e - T_a)
\end{align}
where $C_e$ represents the electronic specific heat
as a function of electronic temperature $T_e$,
$\rho_e$ is the electronic density,
$\kappa_e$ is the electronic thermal conductivity,
$g_p$ is the coupling constant for electron-ion interactions,
and $T_a$ is the ionic temperature \cite{duffy2006, rutherford2007}.
The electronic specific heat was expressed as $C_e = \gamma T_e$,
where $\gamma = \num{4e-9}$ eV/(K$^2$e).
The electronic density was set to $\rho_e = 625$ e/nm$^3$,
and the thermal conductivity was calculated using $\kappa_e = D_e \rho_e C_e$,
where $D_e = 100$ nm$^2$/ps is the thermal diffusion coefficient
\cite{li2017, kolotova2017}.
Although these TTM parameter values have only been used
for pure U or U-5at.\%Mo,
it is assumed that reasonable accuracy can be obtained with these values
for the U-10Mo system as well.

Electronic pressure effects were included in the model
to account for the blast force acting on ions
due to the electronic pressure gradient \cite{chen2006, norman2013}.
Thus, the total force acting on an ion is:
\begin{align}
	\vec{F}_i
		&= -\frac{\partial U}{\partial \vec{r}_i}
		+ \vec{F}_{langevin} - \nabla P_e / n_{ion}
\end{align}
where $\vec{F}_{langevin}$ is the force from the Langevin thermostat
simulating electron-phonon coupling,
$U$ is the potential energy of the system,
$\nabla P_e / n_{ion}$ is the electron blast force,
and $n_{ion}$ is the ion concentration.
The electronic pressure was modeled as $P_e = 0.5 \rho_e C_e T_e$
\cite{norman2013, pisarev2014, kolotova2017}.

\section{Energy loss of fission fragments in U-10Mo}

To understand and quantify re-solution in U-10Mo,
FF behavior must first be investigated.
Fission of \ce{_{92}^{235}U} produces a wide range of isotopes.
To keep computational complexity manageable,
two isotopes, \Y and \I,
were selected as representative light and heavy FFs, respectively.
These isotopes are produced in the fuel via the following nuclear reaction:
\begin{align}
	\ce{
		_0^1n + _{92}^{235}U -> _{39}^{97}Y + _{53}^{136}I + 3 _0^1n + Q
	}
	\label{eq:iso}
\end{align}
While \Y has an initial kinetic energy of approximately $101.3$ MeV,
\I starts with roughly $74.6$ MeV.
These two isotopes were chosen because they correspond
to the two peaks commonly observed in fission product yield distributions,
and they each have a yield of about $0.12$
\cite{setyawan2018, mills1995}.

\begin{figure}[!ht]
	\begin{subfigure}{0.49\textwidth}
		\centering
		\caption{}
		\includegraphics[width=8cm]{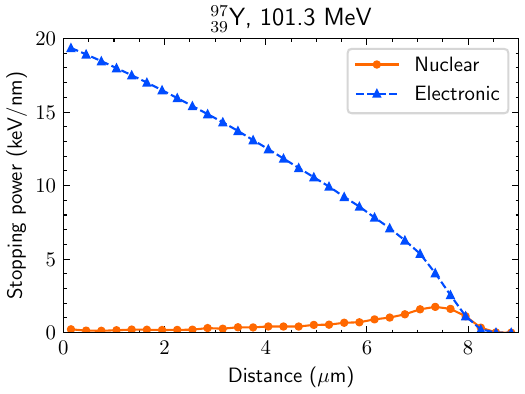}
	\end{subfigure}
	\begin{subfigure}{0.49\textwidth}
		\centering
		\caption{}
		\includegraphics[width=8cm]{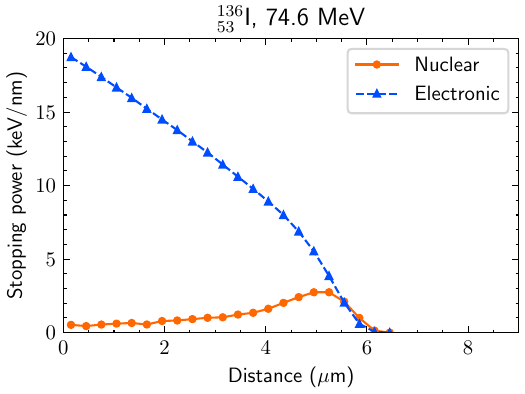}
	\end{subfigure}
	\caption{
		\textbf{Nuclear and electronic stopping powers of FFs.}
		Nuclear and electronic stopping powers of
		(\textbf{a}) light FF \Y with an initial energy of $101.3$ MeV,
		and (\textbf{b}) heavy FF \I with an initial energy of $74.6$ MeV
		as a function of distance in U-10Mo.
		Data were obtained from $2,000$ independent BCA simulations
		for each FF using RustBCA.
	}
	\label{fig:stopping}
\end{figure}

To evaluate the energy loss of these FFs in U-10Mo,
BCA simulations were conducted using RustBCA's \texttt{0D} geometry option.
In these simulations, U-10Mo was modeled as an infinite medium
extending from $0$ to $\infty$ in the $x$ direction,
and from $-\infty$ to $\infty $ in the $y$ and $z$ directions.
The FFs were injected at the origin
with an initial velocity oriented along the $x$ axis.
For each FF, $2,000$ independent ion-material simulations were performed.
From these simulations,
FF positions and velocities from each binary collision
were recorded, then processed and binned
to obtain electronic and nuclear stopping power profiles,
as illustrated in Figure \ref{fig:stopping}.
At the beginning of the FF trajectories,
the electronic stopping power is slightly less than $20$ keV/nm for both FFs
and decreases almost linearly with distance.
In contrast, the nuclear stopping power remains low
for the majority of the FF trajectory,
peaking only at the very end of the path
as the FF reaches energies where nuclear collisions become more probable.
The nuclear energy loss for \Y accounts for
approximately $5\%$ of its total initial energy,
whereas for \I, the nuclear energy loss is around $10\%$.

Using insights derived from the stopping power profiles,
two complementary methods were employed
to investigate gas bubble re-solution in U-10Mo.
To analyze the effect of electronic stopping,
the TTM was utilized to simulate the energy transfer
between the electronic subsystem and the ionic subsystem
(discussed further in Section \ref{sec:elec}).
To capture the effect of nuclear stopping on re-solution,
the interactions among FFs, U-10Mo, and Xe gas bubbles were simulated
using the newly implemented \texttt{SPHEREINCUBOID} geometry in RustBCA,
as detailed in Section \ref{sec:nuke}.

\section{Re-solution due to electronic stopping}
\label{sec:elec}

To simulate thermal spikes, a simulation cell with dimensions
$120 \alpha_0 \times 120 \alpha_0 \times 50 \alpha_0$
($\alpha_0 = 0.343$ nm is the lattice parameter of U-10Mo at $400$ K
\cite{phillips2010})
was created with periodic boundary conditions in all directions.
A random distribution of U and Mo atoms in a bcc lattice
was first generated to achieve a Mo concentration of $22$ at.\%.
Using this configuration,
a spherical gas bubble with a radius of $2$ nm was created
by removing U and Mo atoms and depositing Xe atoms inside the void.
The Xe gas atoms were introduced at a Xe/vacancy ratio of $0.2$,
resulting in approximately $330$ Xe atoms within the bubble.
The system was equilibrated at $400$ K and $0$ bar
by using an NPT ensemble for $10$ ps, with a timestep size of $1$ fs.
The Nos\'e-Hoover thermostat and barostat
regulated the temperature and pressure during equilibration.
The bubble size and Xe/vacancy ratio at the prescribed temperature and pressure
follow the results from the equation of state of Beeler et al.
\cite{beeler2020improved}.

To solve heat transfer related to the electronic subsystem,
electronic cells with dimensions
$2 \alpha_0 \times 2 \alpha_0 \times 50 \alpha_0$
were defined across the whole simulation cell,
with periodic boundary conditions applied in all directions.
A thermal spike was introduced
by initializing the electronic temperature profile in these cells according to:
\begin{align}
	T_e &= T_{init} + T_{spike} \exp\left( -r / R \right)
\end{align}
where $r$ is the radial distance from the thermal spike axis,
$R = 10 \alpha_0$, and $T_{init} = 400$ K.
The thermal spike axis was aligned with the shortest dimension of the supercell.
The values of $T_{spike}$ were chosen
according to the desired electronic stopping power.
The simulations were then performed using an NVE ensemble
with a canonical sampling thermostat \cite{bussi2007}
applied at the edges of the simulation box to serve as a heat sink.
Only those edges that did not intersect the thermal spike axis
were included in the heat sink region.
A variable timestep size was implemented
such that the maximum displacement of any atom
between two successive timesteps was less than or equal to $0.001$ nm.
$100,000$ timesteps were performed for each simulation,
yielding $80$--$90$ ps of total simulation time.

\begin{figure}[!ht]
	\centering
	\begin{subfigure}{0.49\textwidth}
		\centering
		\caption{}
		\includegraphics[width=8cm]{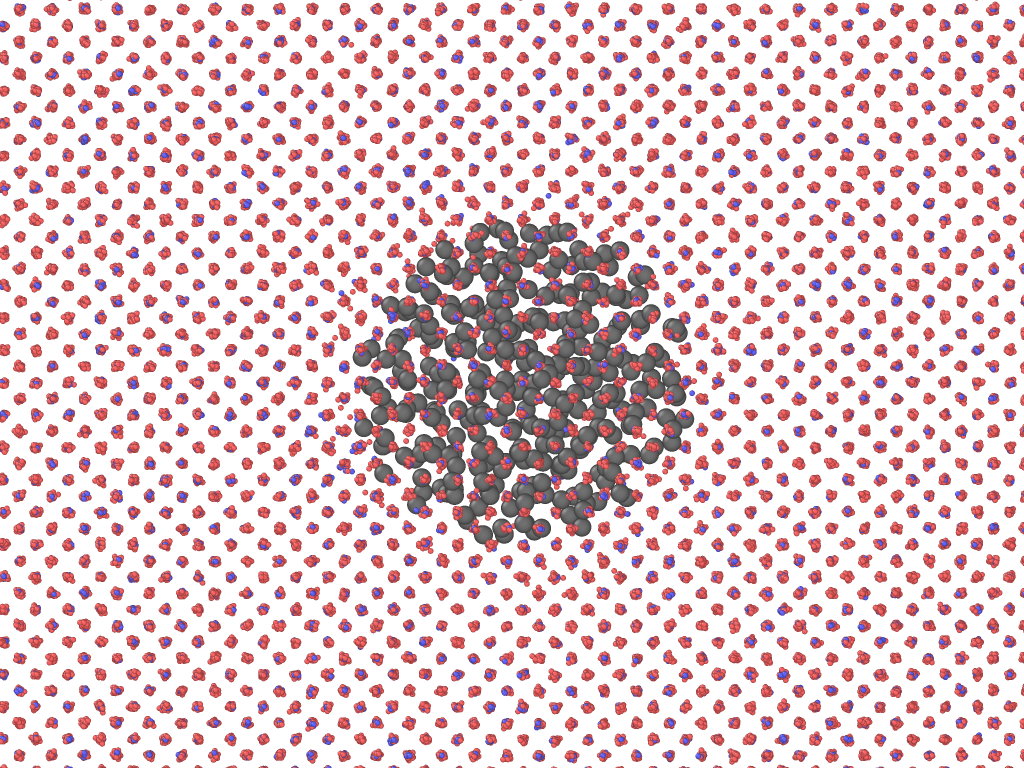}
	\end{subfigure}
	\begin{subfigure}{0.49\textwidth}
		\centering
		\caption{}
		\includegraphics[width=8cm]{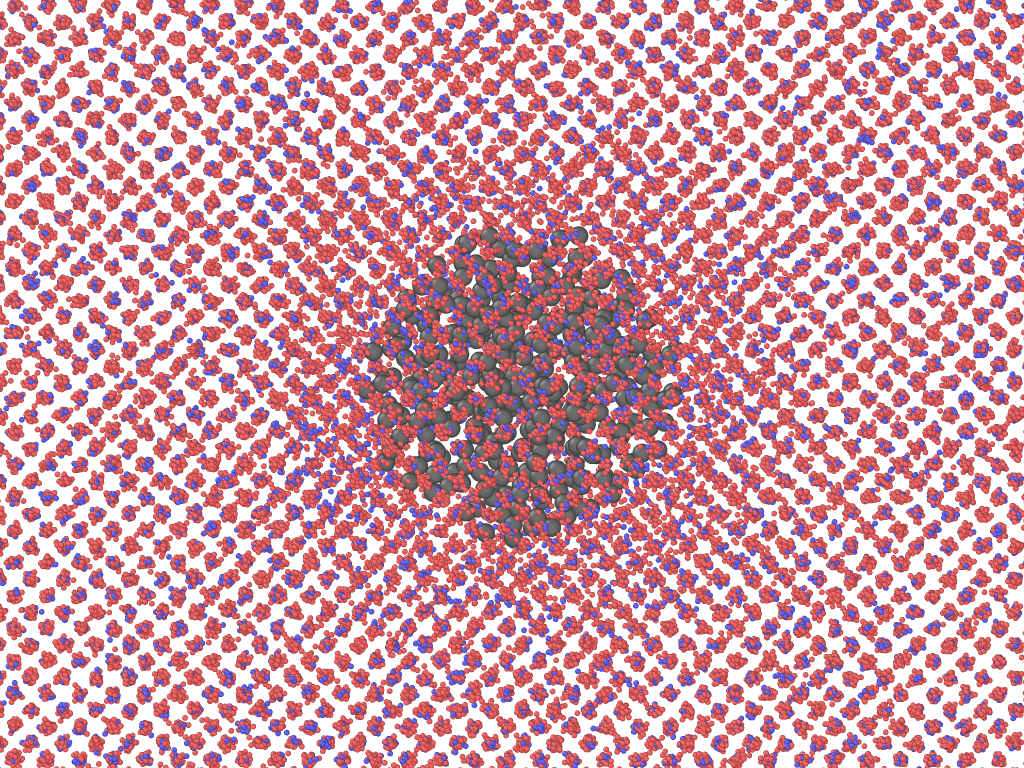}
	\end{subfigure}
	\begin{subfigure}{0.49\textwidth}
		\centering
		\caption{}
		\includegraphics[width=8cm]{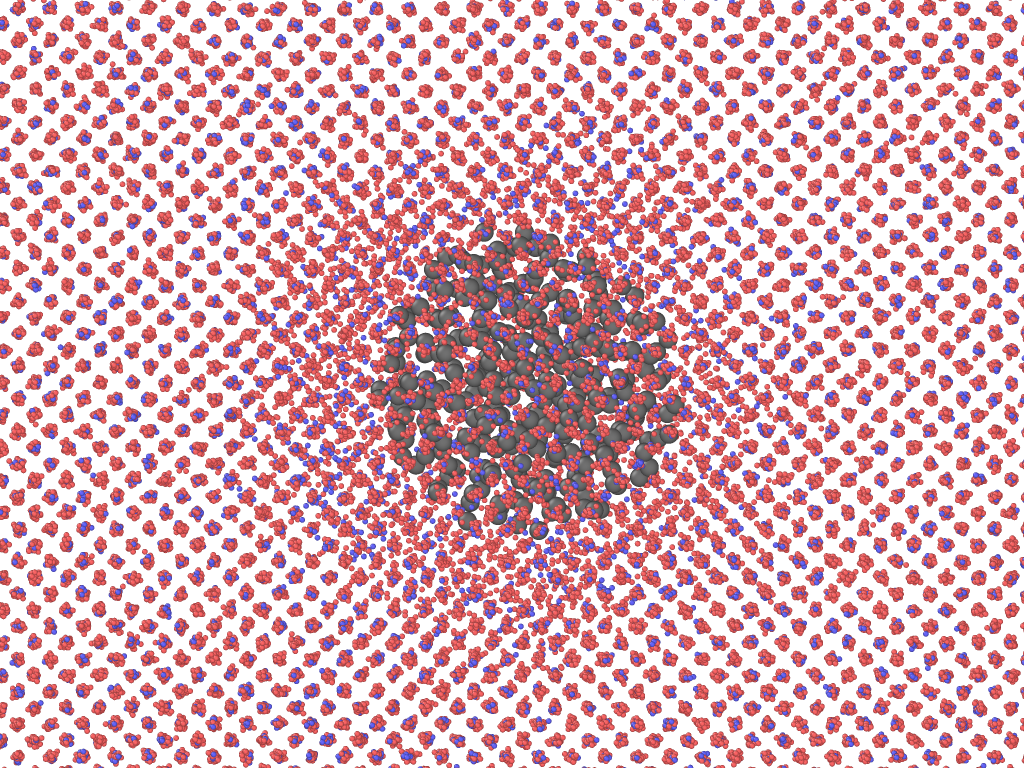}
	\end{subfigure}
	\begin{subfigure}{0.49\textwidth}
		\centering
		\caption{}
		\includegraphics[width=8cm]{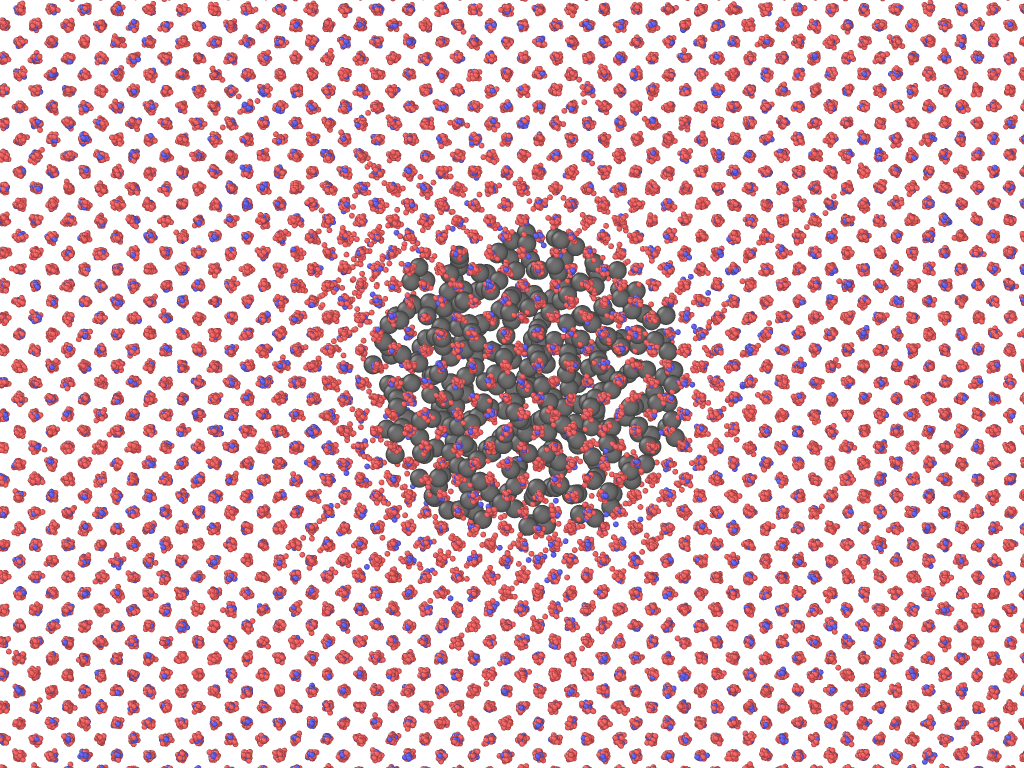}
	\end{subfigure}
	\caption{
		\textbf{Temporal evolution of a thermal spike simulation in U-10Mo.}
		Snapshots of a $30$ keV/nm thermal spike simulation
		at (\textbf{a}) $0$ ps, (\textbf{b}) $2.4$ ps, (\textbf{c}) $10.1$ ps,
		and (\textbf{d}) $29.1$ ps.
		Xe gas atoms are represented in black, U in red, and Mo in blue.
		The local ionic temperature rises rapidly for $1.5$ ps
		before beginning to cool.
		No Xe atom was observed to escape the gas bubble.
		The images were rendered using OVITO.
	}
	\label{fig:ttm}
\end{figure}

Simulations were performed with three values of $T_{spike}$
($28,000$ K, $31,500$ K, and $34,500$ K),
corresponding to electronic stopping powers
ranging from $20$ keV/nm to $30$ keV/nm.
No re-solution events were observed in any of these simulations.
Figure \ref{fig:ttm} presents a few snapshots
of a $30$ keV/nm thermal spike simulation as visualized in OVITO \cite{ovito}.
The local ionic subsystem temperature rises rapidly for about $1.5$ ps
following the initiation of the thermal spike,
reaching its peak before beginning to cool.
After $30$ ps, only a few defects are visible within the system.
The local temperature gradually returns to the initial equilibrium temperature
over approximately $60$ ps.

Kolotova et al. reported threshold electronic stopping powers
for defect formation and melting in U-5at.\%Mo at various temperatures
\cite{kolotova2017}.
At $400$ K, the threshold stopping powers for defect formation and melting
were found to be approximately $22$ keV/nm and $26$ keV/nm, respectively.
Given that 1) no gas bubble re-solution was observed
in the MD simulations of $30$ keV/nm thermal spikes,
2) peak stopping power of FFs in U-10Mo is approximately $20$ keV/nm
(Figure \ref{fig:stopping}),
and 3) threshold stopping power for defect formation exceeds $20$ keV/nm,
it is highly unlikely that gas bubble re-solution can occur in U-10Mo
through the heterogeneous mechanism.
This behavior is expected since U-10Mo is a metallic system
and displays a relatively high thermal conductivity,
but it has been explicitly confirmed here for the first time.

\section{Re-solution due to nuclear stopping}
\label{sec:nuke}

\subsection{Model for re-solution calculation}

\begin{figure}[!ht]
	\begin{subfigure}{\textwidth}
		\centering
		\caption{}
		\includegraphics[height=5cm]{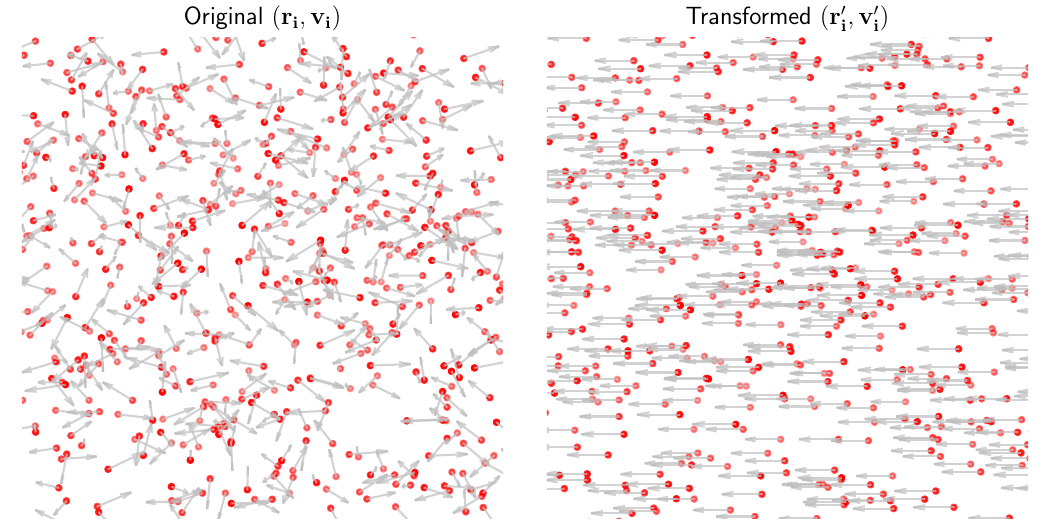}
	\end{subfigure}
	\begin{subfigure}{\textwidth}
		\centering
		\caption{}
		\includegraphics[height=7cm]{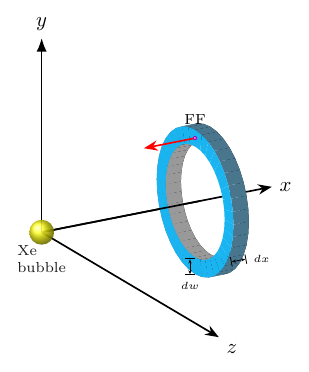}
	\end{subfigure}
	\caption{
		\textbf{Coordinate system for the calculation of the re-solution rate.}
		\textbf{a} Rotation of fission event positions and velocities
		around the origin such that all velocities point to the $-x$ direction.
		Red dots represent positions and gray arrows represent velocities.
		\textbf{b} A coordinate system in which a Xe gas bubble is at the origin
		and FFs are all pointing toward the $-x$ direction.
		The radial coordinate $w = \sqrt{y^2 + z^2}$
		represents the perpendicular distance between
		the initial FF trajectory and the bubble center.
	}
	\label{fig:coord}
\end{figure}

The re-solution rate can be defined as the probability
of a Xe atom escaping a gas bubble and entering the surrounding fuel matrix,
per unit time.
To capture the overall re-solution behavior in the material,
the contributions from all the FFs,
originating at various distances from the bubble
and oriented toward random directions,
must be integrated over the entire volume of interest.
Consider a 3D coordinate system
in which the gas bubble is located at the origin.
To simplify the integration,
FF points of origin can be rotated around the coordinate system origin
such that their initial velocities align in the same direction.
Since FF generation in the material is uniform and isotropic,
this rotational transformation results in
a uniform distribution of unidirectional FFs,
as shown in Figure \ref{fig:coord}a.
The axial coordinate $x$ is then defined as
parallel to the FF velocities (which are now directed toward $-x$),
whereas the radial coordinate $w := \sqrt{y^2+z^2}$ is defined as
perpendicular to the velocities (Figure \ref{fig:coord}b).
If the fission rate is denoted as $\dot{F}$,
the number of fission events per second
in an infinitesimal volume $dV = 2 \pi w \: dw  \: dx$ would be $\dot{F} dV$.
Also, we assume that when a FF isotope $k$ originating at $(x, w)$
interacts with a Xe gas bubble at the origin,
it re-solves a fraction, $\xi_k \equiv \xi_k(x, w)$, of the Xe atoms.
All the $k$ isotopes from fission events in $dV$ thus contribute
$\xi_k(x, w) \dot{F} dV$ to the total re-solution of the bubble.
The re-solution rate $b$ can then be expressed as:
\begin{align}
	b &= \sum_{k = Y, I} \int_V \xi_k(x, w) \dot{F} dV \label{eq:b} \\
	  &= \dot{F} \sum_{k = Y, I} \int_V \xi_k(x, w) dV
	   = \dot{F} \left( \int_V \xi_Y(x, w) dV + \int_V \xi_I(x, w) dV \right) \\
	  &= \dot{F} \sum_{k = Y, I} \int_{x=0}^{\infty} \int_{w=0}^{\infty}
		\xi_k(x, w) 2 \pi w dw dx
\end{align}
where it is assumed that all fission events produce \Y and \I,
as per Equation \ref{eq:iso}.

\subsection{Reference fission fragment simulations}

The most straightforward method to calculate $\xi_k(x, w)$ involves
simulating a FF and a Xe gas bubble for a specific $(x, w)$ value.
However, performing these simulations for all required $(x, w)$ values
is computationally expensive and often yields statistically unreliable results.
As the distance between the origin of a given FF and a Xe gas bubble increases,
the probability of interaction between them decreases significantly.
The interaction probability is also influenced by the size of the gas bubble:
smaller bubbles exhibit a lower probability of interaction.
Therefore, for certain bubble sizes and $(x, w)$ values,
even conducting hundreds of thousands of BCA simulations
may result in only a few interactions,
making most of these simulations an inefficient use of computational resources.

\begin{figure}[!ht]
	\centering
	\begin{subfigure}{0.49\textwidth}
		\centering
		\caption{}
		\includegraphics[height=5cm]{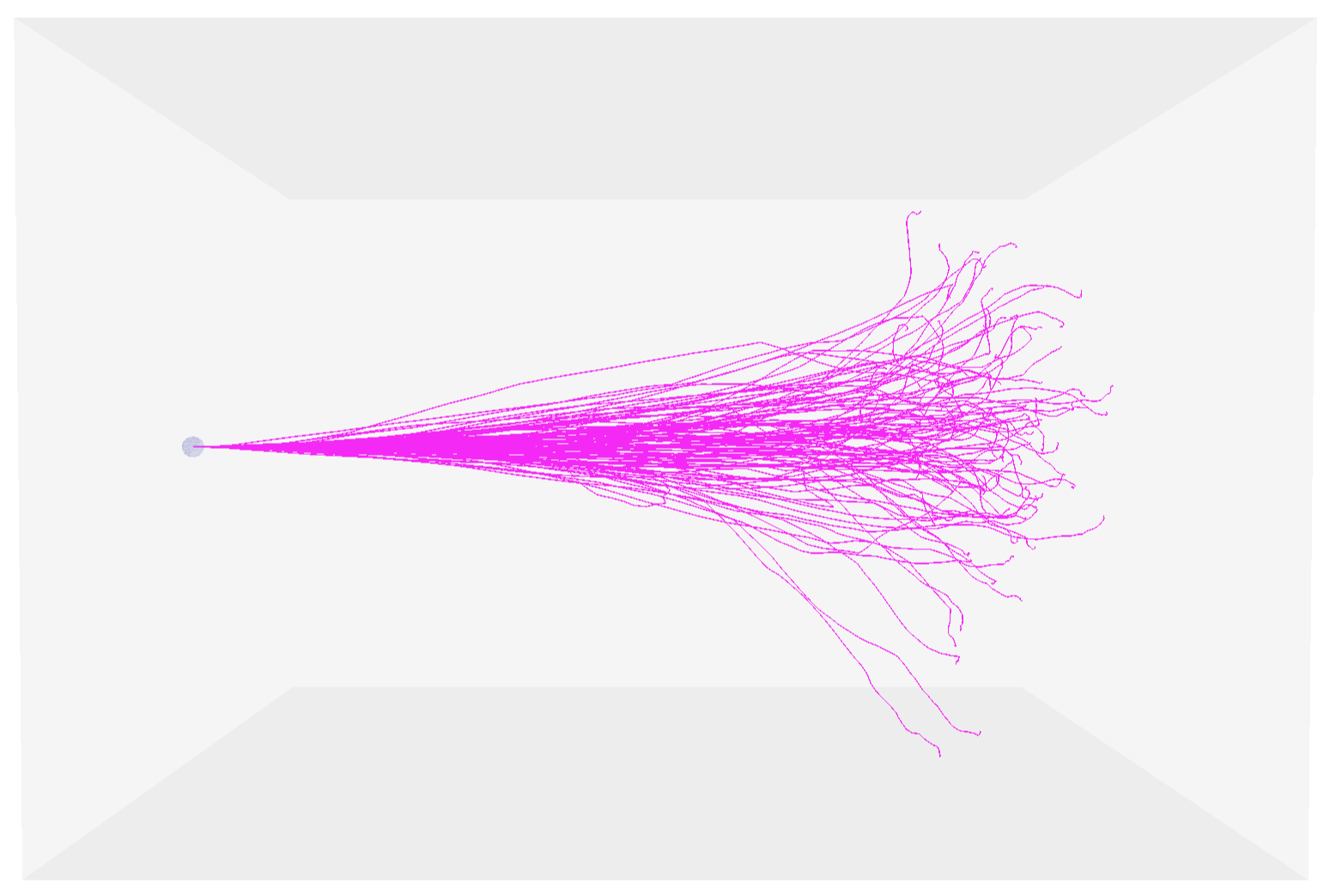}
	\end{subfigure}
	\begin{subfigure}{0.49\textwidth}
		\centering
		\caption{}
		\includegraphics[height=5cm]{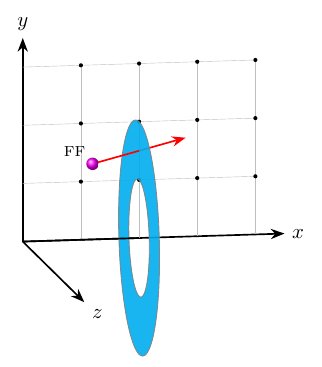}
	\end{subfigure}
	\caption{
		\textbf{FF trajectory discretization scheme.}
		\textbf{a} Trajectories of $100$ simulated \Y ions in U-10Mo,
		starting from the origin and directed along the $x$ axis.
		\textbf{b} Schematic of the annular surface discretization scheme
		used to collect data across the volume.
		The discretization allows for the extraction of three key properties
		at each $(x, w)$ coordinate:
		the probability of
		a FF passing through a unit surface area centered at $(x, w)$,
		its average incidence energy,
		and its average incidence angle relative to the $x$ axis.
		The simulation volume is discretized using a two-dimensional grid
		with uniform spatial intervals of $\Delta x = \Delta w = 50$ nm.
	}
	\label{fig:fftrack}
\end{figure}

An alternative to the brute-force approach involves analyzing
the behavior of FFs in the fuel matrix
without the presence of any gas bubbles first.
If the probability of a FF
passing through a specific point in the fuel with a given energy is known,
assessing re-solution behavior from local FF--bubble interactions
at that exact point then becomes straightforward.
Thus, FF simulations in U-10Mo were performed
to obtain three key properties for each $(x, w)$ coordinate:
the probability of
a FF passing through a unit surface area centered at $(x, w)$,
its average incidence energy,
and its average incidence angle relative to the $x$ axis.
In these reference simulations, the FF started from the origin
and was directed toward the $x$ axis within a U-10Mo matrix,
as displayed in Figure \ref{fig:fftrack}a.
It is important to note that this setup is inverted
from the configuration shown in Figure \ref{fig:coord}b,
where the Xe gas bubble is positioned at the origin.

To collect data from the simulations,
an annular surface discretization scheme,
as illustrated in Figure \ref{fig:fftrack}b, was used.
The discretized surface elements were associated with specific $(x, w)$ values,
with the spacing between successive surface elements
in the $x$ or $w$ direction being set to $\Delta x = \Delta w = 50$ nm.
This level of discretization was sufficient
to produce smooth profiles for the FF properties.
To verify the convergence of these FF profiles,
six points were selected for both \Y and \I.
Simulations were conducted in batches of $1000$ ions.
The simulations were terminated
when the relative changes in probability, energy, and angle
between two consecutive batches of simulations fell below $0.001$
at each of the six selected points.
The following $(x, w)$ coordinates, measured in $\mu$m,
were used for convergence:
$(3, 0)$, $(5, 0)$, $(7, 0)$,
$(4.5, 0.5)$, $(6.5, 0.5)$, and $(6, 1)$ for \Y,
and $(2, 0)$, $(3.5, 0)$, $(5, 0)$,
$(3, 0.5)$, $(4.5, 0.5)$, and $(4, 1)$ for \I.
The \Y profiles converged after $30,000$ simulations,
whereas the \I profiles required $40,000$ simulations to achieve convergence.

\begin{figure}[!ht]
	\centering
	\begin{subfigure}{0.49\textwidth}
		\centering
		\caption{}
		\includegraphics
			[width=8cm, trim={0.8cm 0 1.5cm 0.4cm}, clip]
			{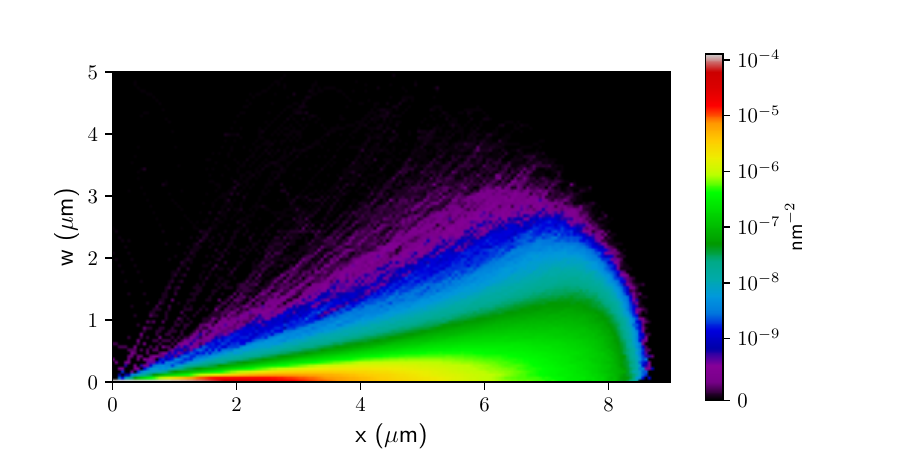}
	\end{subfigure}
	\begin{subfigure}{0.49\textwidth}
		\centering
		\caption{}
		\includegraphics
			[width=8cm, trim={0.8cm 0 1.5cm 0.4cm}, clip]
			{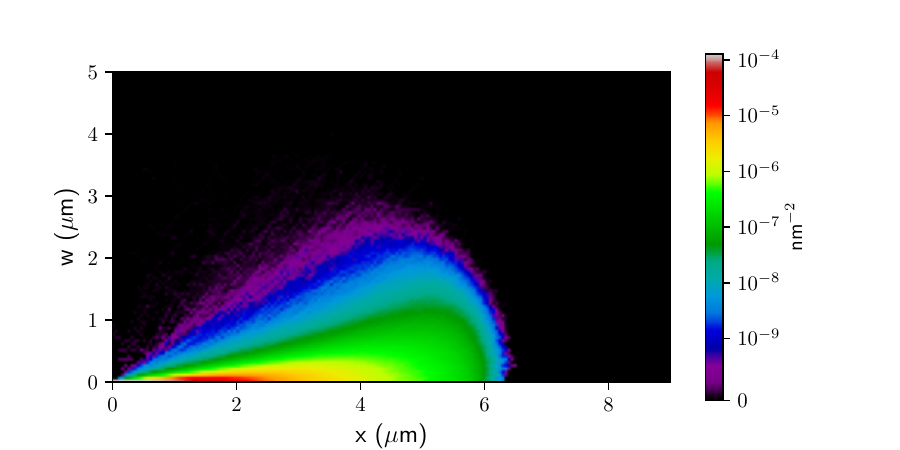}
	\end{subfigure}
	\begin{subfigure}{0.49\textwidth}
		\centering
		\caption{}
		\includegraphics
			[width=8cm, trim={0.8cm 0 1.5cm 0.4cm}, clip]
			{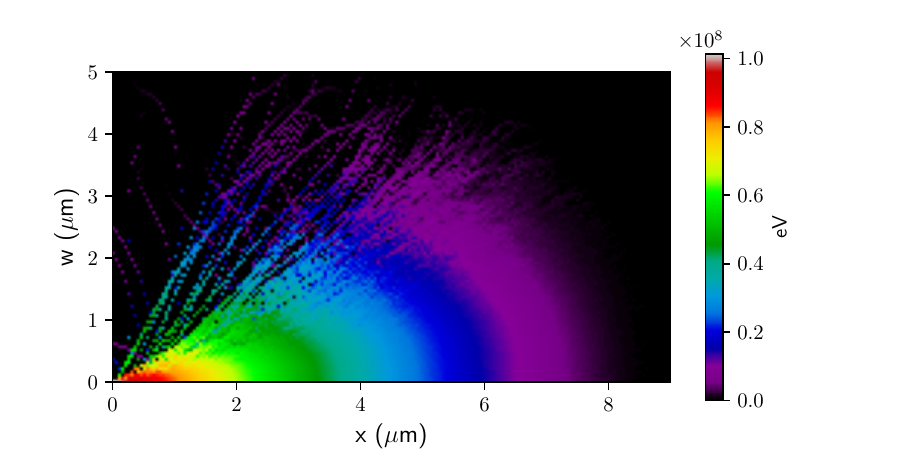}
	\end{subfigure} \begin{subfigure}{0.49\textwidth}
		\centering
		\caption{}
		\includegraphics
			[width=8cm, trim={0.8cm 0 1.5cm 0.4cm}, clip]
			{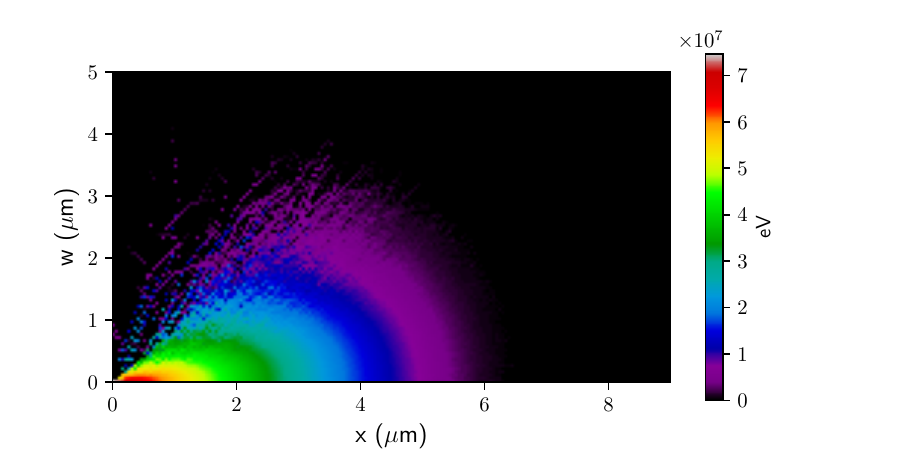}
	\end{subfigure}
	\begin{subfigure}{0.49\textwidth}
		\centering
		\caption{}
		\includegraphics
			[width=8cm, trim={0.8cm 0 1.5cm 0.4cm}, clip]
			{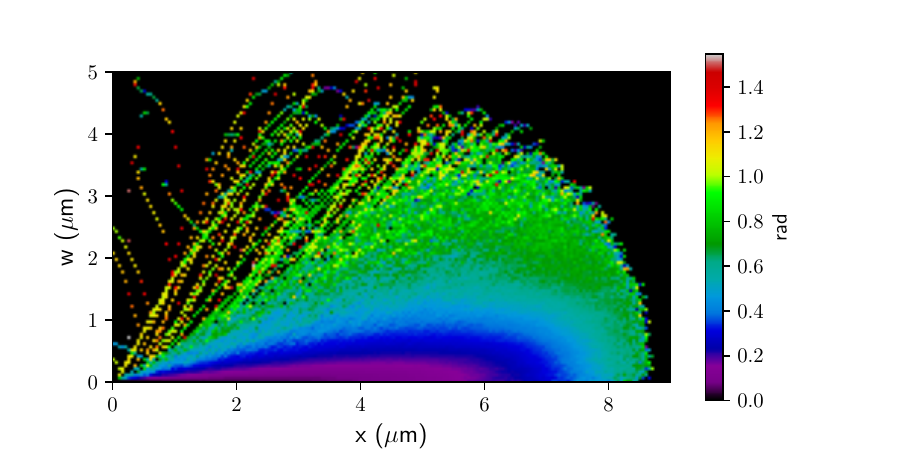}
	\end{subfigure}
	\begin{subfigure}{0.49\textwidth}
		\centering
		\caption{}
		\includegraphics
			[width=8cm, trim={0.8cm 0 1.5cm 0.4cm}, clip]
			{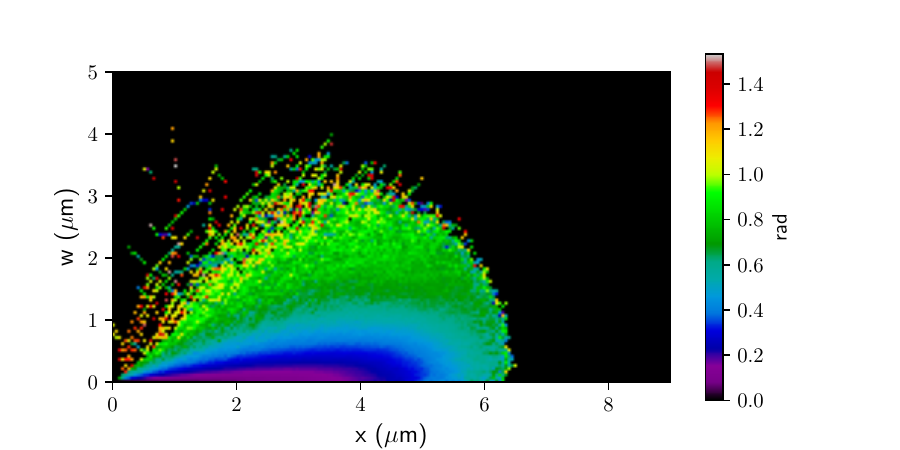}
	\end{subfigure}
	\caption{
		\textbf{Spatial distribution of FF incidence properties in U-10Mo.}
		Calculated FF properties in U-10Mo as a function of coordinates $(x, w)$
		for FFs initiated at the origin and directed along the $x$ axis.
		Data represent the converged results
		from $30,000$ simulations of \Y and $40,000$ simulations of \I.
		\textbf{a, b}
		Ion incidence probability per unit surface area centered at $(x, w)$
		for (\textbf{a}) \Y and (\textbf{b}) \I.
		The probability profiles show a plume-like broadening
		as FFs travel further from the origin.
		\textbf{c, d}
		Average incidence energy for (\textbf{c}) \Y and (\textbf{d}) \I,
		showing a predominantly linear energy loss with distance.
		\textbf{e, f}
		Average incidence angle relative to the $x$ axis for (\textbf{e}) \Y
		and (\textbf{f}) \I.
		Discrete paths visible in the low-probability regions
		of the energy and angle plots represent rare scattering events.
	}
	\label{fig:ref}
\end{figure}

Figure \ref{fig:ref} presents the results
from the simulations and subsequent discretization.
The incidence probability per unit area
is displayed in Figures \ref{fig:ref}a and \ref{fig:ref}b.
The probability profiles broaden as $x$ increases,
producing a plume-like pattern.
Figures \ref{fig:ref}c and \ref{fig:ref}d
illustrate the incidence ion energies.
The circular pattern clearly demonstrates how the ions lose energy
as they travel farther from the origin,
with the energy loss as a function of distance being predominantly linear.
Lastly, Figures \ref{fig:ref}e and \ref{fig:ref}f display
the incidence angle of ions with respect to the $x$ axis.
As expected, ions closer to $w = 0$ have a low incidence angle,
while ions farther away show higher incidence angles.
In both the incidence energy and incidence angle plots,
a few discrete FF paths are visible in the top-left region.
These are highly unlikely occurrences,
as is evident from the probability figures.
The ion profiles also reveal the ranges of \Y and \I ions in U-10Mo
to be approximately $8.5$ $\mu$m and $6.5$ $\mu$m, respectively.
The observed ranges, shown in Figure \ref{fig:ref},
are consistent with the data presented in Figure \ref{fig:stopping}.

\subsection{Fission fragment interactions with Xe gas bubbles}

\begin{figure}[!ht]
	\centering
	\includegraphics[width=8cm]{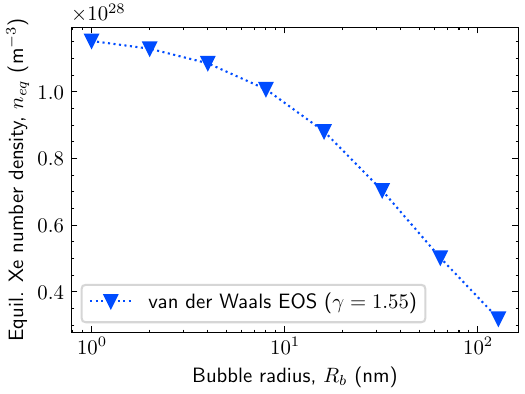}
	\caption{
		\textbf{Equilibrium Xe number density as a function of bubble radius.}
		The equilibrium Xe number density within the gas bubbles was calculated
		using the van der Waals EOS.
		The equilibrium bubble pressure used in the EOS was derived
		from the Young-Laplace equation,
		utilizing a surface energy of $\gamma = 1.55$ J/m$^2$.
	}
	\label{fig:vdw}
\end{figure}

With the reference FF properties available,
the next step involves simulating local FF--bubble interactions
to quantify the gas bubble re-solution rate.
For these simulations, equilibrium Xe number densities were used
for bubbles of all sizes.
The equilibrium number densities were calculated
using the van der Waals equation of state (EOS),
as described in \cite{olander1975}:
\begin{align}
	n &= \left( B + \frac{kT}{p} \right)^{-1} \\
	n_{eq} &= \left( B + \frac{kT}{p_{eq}} \right)^{-1}
	        = \left( B + \frac{kT R_b}{2 \gamma} \right)^{-1}
\end{align}
The equilibrium bubble pressure $p_{eq}$ utilized in the EOS
was derived from the Young-Laplace equation:
\begin{align}
	p_{eq} &= \frac{2 \gamma}{R_b}
\end{align}
where $R_b$ is the bubble radius
and $\gamma = 1.55$ J/m$^2$ is the surface energy in U-10Mo \cite{beelerADP}.
The resulting equilibrium number densities
are plotted as a function of bubble radius in Figure \ref{fig:vdw}.
Notably, the van der Waals EOS predicts
a plateau in the equilibrium number density for smaller bubbles.

\begin{figure}[!ht]
	\centering
	\includegraphics[width=8cm]{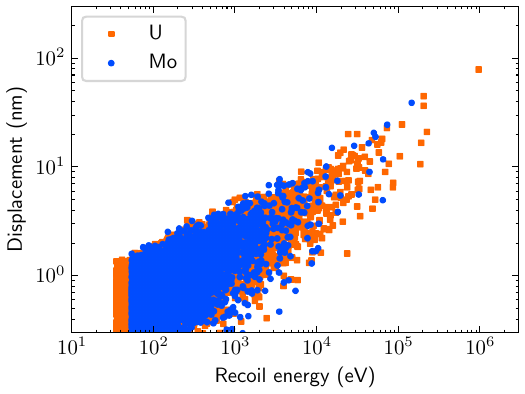}
	\caption{
		\textbf{Recoil displacement as a function of recoil energy in U-10Mo.}
		The scatter plot displays the data
		collected from five independent BCA simulations of \Y FFs.
		The maximum energy transferred to a recoil is approximately $1$ MeV,
		resulting in a displacement of roughly $100$ nm.
		This displacement ($\delta = 100$ nm) was used to establish
		a spatial cut-off for subsequent simulations of FF--bubble interactions,
		assuming the recoils generated beyond this distance from a gas bubble surface
		are unlikely to interact with it.
	}
	\label{fig:recoil}
\end{figure}

\begin{figure}[!ht]
	\centering
	\includegraphics[width=7.5cm]{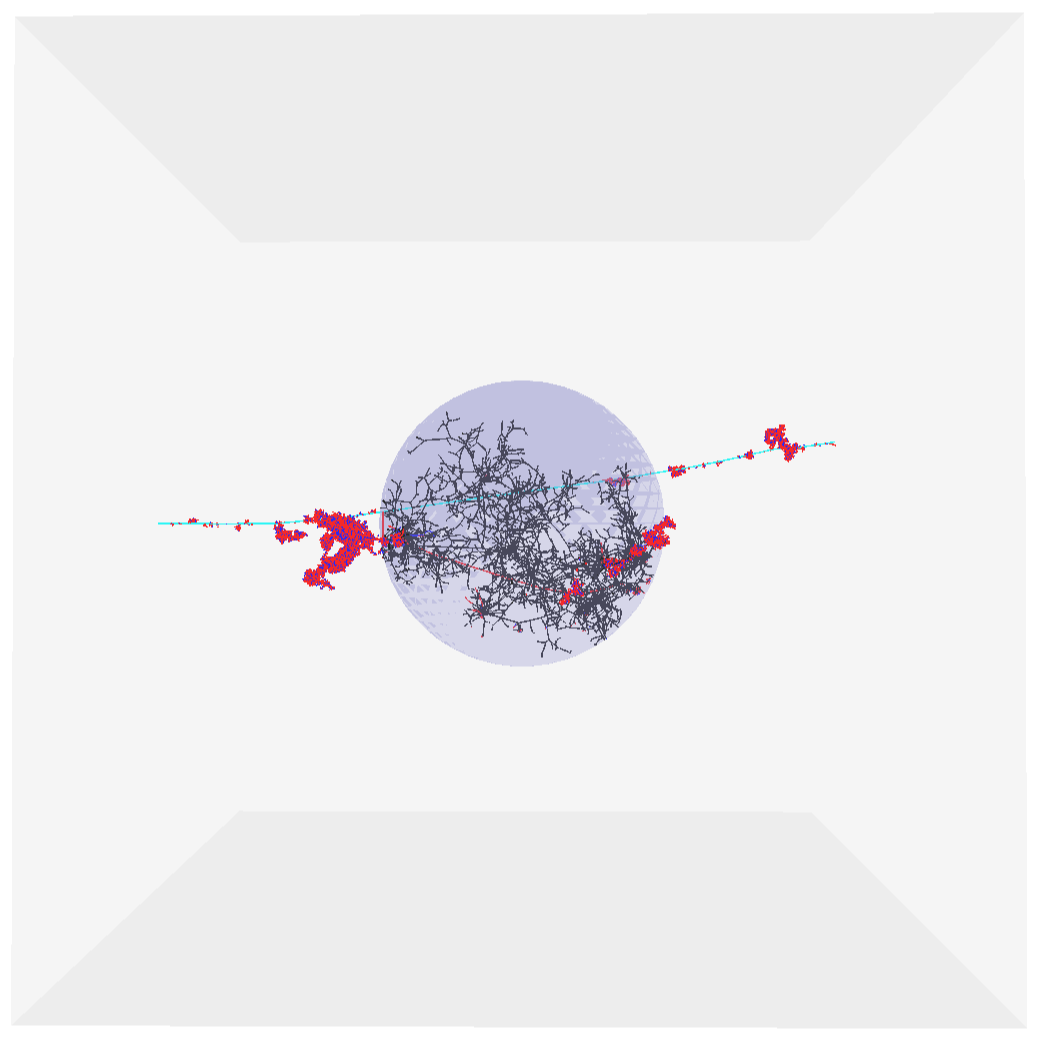}
	\caption{
		\textbf{Recoil trajectories from FF--bubble simulations.}
		Visualization of an individual BCA simulation
		where a $5$ MeV \Y FF is incident on a Xe gas bubble with a $64$ nm radius.
		Red, blue, black, and cyan lines represent U, Mo, Xe, and \Y, respectively.
		In this specific interaction, $3,823$ Xe recoils were generated,
		with $155$ Xe atoms being displaced outside the bubble surface
		and $9$ atoms being re-solved successfully
		(defined as being displaced $\geq 1$ nm away from the bubble surface).
		The image was rendered using VisPy.
	}
	\label{fig:ffbub}
\end{figure}

FF--bubble interactions should be simulated in a manner
that accounts for all relevant recoils.
To this end, five BCA simulations of \Y in U-10Mo
were analyzed to evaluate recoil behavior.
Figure \ref{fig:recoil} presents a scatter plot
of the recoil displacement as a function of recoil energy.
The maximum energy transferred to a recoil atom was about 1 MeV,
with the recoil displacement being around $100$ nm.
While it is theoretically possible to transfer more than 1 MeV
to a recoil in a head-on collision,
such events are extremely rare.
Therefore, it is reasonable to assume that
recoils generated by FFs more than $\delta = 100$ nm away from a gas bubble
will not interact with that bubble.
In the FF--bubble BCA simulations, the FFs were offset
by a distance $D = R_b + \delta$ along the $x$ axis from the bubble center.
Recoil trajectories from one such simulation are visualized
in Figure \ref{fig:ffbub} using VisPy \cite{vispy}.

\begin{figure}[!ht]
	\begin{subfigure}{0.49\textwidth}
		\centering
		\caption{}
		\includegraphics[width=8cm]{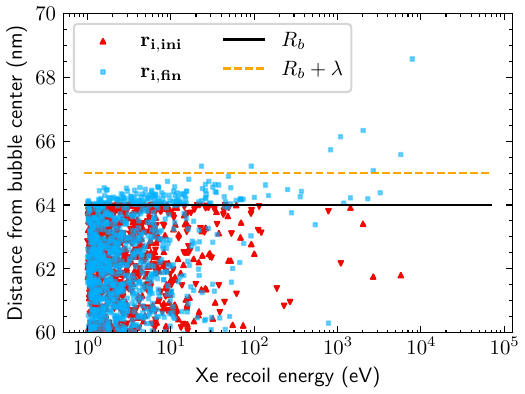}
	\end{subfigure}
	\begin{subfigure}{0.49\textwidth}
		\centering
		\caption{}
		\includegraphics[width=8cm]{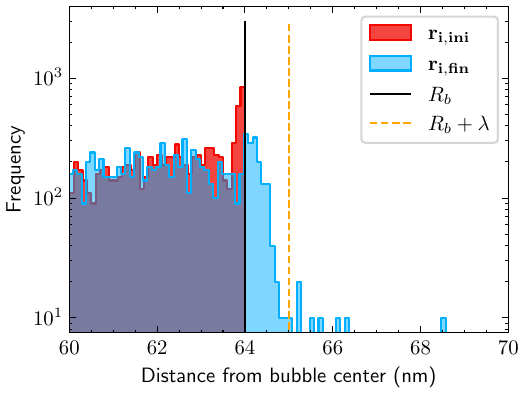}
	\end{subfigure}
	\caption{
		\textbf{Analysis of Xe recoils in a FF--bubble interaction.}
		The data were obtained from a simulation involving a $5$ MeV \Y
		incident on a Xe bubble with a radius of $R_b = 64$ nm.
		\textbf{a} Initial ($\mathbf{r_{i,ini}}$)
		and final ($\mathbf{r_{i,fin}}$) positions of Xe recoils
		relative to the bubble center plotted as a function of recoil energy.
		The plot indicates
		a minimum threshold energy of $E_{min} \approx 25$ eV,
		corresponding to the re-solution cut-off distance of $\lambda = 1$ nm.
		\textbf{b} Histogram showing the distribution
		of the initial and final positions of Xe recoils.
		The majority of the recoils originate near the bubble surface
		and are displaced to positions just outside the bubble surface.
		In both graphs, the distance from the bubble center is truncated
		to highlight the recoil behavior at the bubble--matrix interface.
	}
	\label{fig:xeres}
\end{figure}

Next, we determine which Xe recoils are re-solved.
In this work, Xe atoms that end up at least $\lambda = 1$ nm away
from the bubble surface are considered re-solved.
In reality, the bubble surface is not a static structure
due to the thermal fluctuations of the Xe atoms comprising it.
As a result, a Xe atom that is just outside the surface is highly likely
to return to the bubble within a short period.
Thus, a finite annular region outside the bubble surface
must be cleared by a Xe atom in order to be considered fully re-solved.
Our choice of $\lambda$ was informed
by the existing literature on re-solution simulations
\cite{schwen2009md, govers2012, setyawan2018}.
Figure \ref{fig:xeres} depicts the displacements of Xe recoils
from an example simulation of a FF interacting with a bubble.
From a plot of the recoil displacement as a function of recoil energy,
such as Figure \ref{fig:xeres}a,
one can determine
the minimum threshold energy $E_{min}$ required for re-solution,
as $\lambda$ and $E_{min}$ are interconnected.
While some re-solution studies have subjectively chosen $\lambda$,
others have focused on $E_{min}$  as the criterion
\cite{ronchi1986, matthews2015}.
Based on our simulations,
we found $E_{min}$ to be about $25$ eV for $\lambda=1$ nm.
Additionally, Figure \ref{fig:xeres}b shows that
most Xe recoils originate close to the bubble surface
and end up just outside it.

\begin{figure}[!ht]
	\centering
	\begin{subfigure}{0.49\textwidth}
		\centering
		\caption{}
		\includegraphics[width=8cm]{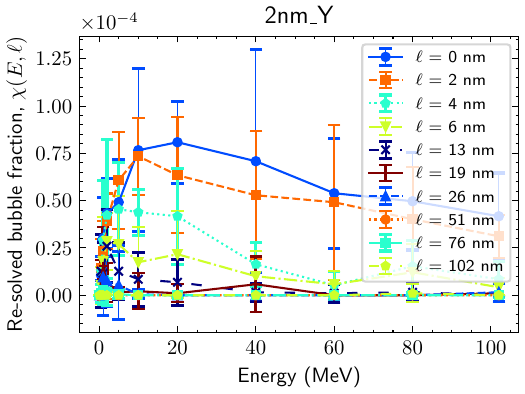}
	\end{subfigure}
	\begin{subfigure}{0.49\textwidth}
		\centering
		\caption{}
		\includegraphics[width=8cm]{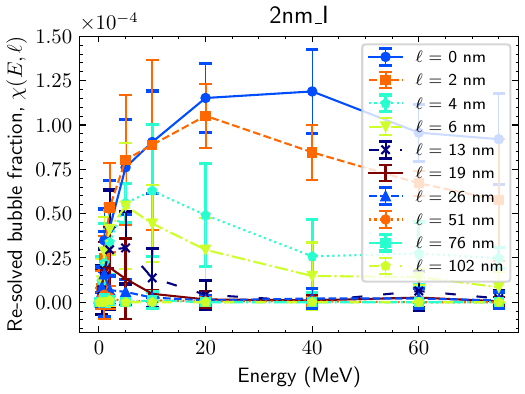}
	\end{subfigure}
	\begin{subfigure}{0.49\textwidth}
		\centering
		\caption{}
		\includegraphics[width=8cm]{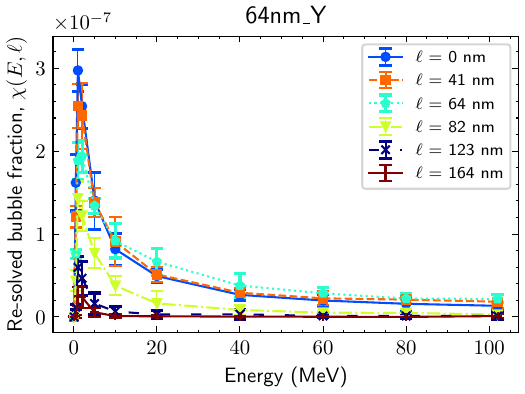}
	\end{subfigure}
	\begin{subfigure}{0.49\textwidth}
		\centering
		\caption{}
		\includegraphics[width=8cm]{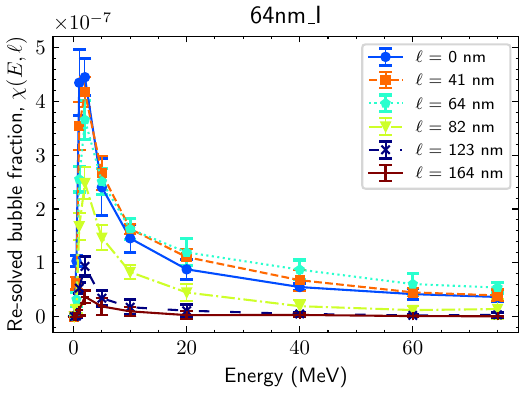}
	\end{subfigure}
	\caption{
		\textbf{Fraction of Xe atoms re-solved from gas bubbles ($\chi$)
		as a function of FF energy ($E$) and off-center distance ($\ell$).}
		The bubble is centered at the origin $(x, w) = (0, 0)$,
		and the FF is initiated at $(x, w) = (-(R_b + \delta), \ell)$
		with an energy $E$ and a direction along the $x$ axis.
		$R_b$ is the bubble radius and $\delta$ is the maximum recoil distance.
		\textbf{a, b} $\chi(E, \ell)$ for a bubble with a $2$ nm radius
		interacting with incident (\textbf{a}) \Y and (\textbf{b}) \I.
		\textbf{c, d} $\chi(E, \ell)$ for a bubble with a $64$ nm radius
		interacting with incident (\textbf{c}) \Y and (\textbf{d}) \I.
		Error bars represent $2 \sigma$ deviations from the mean
		across $5,000$ independent BCA simulations per configuration.
	}
	\label{fig:chi}
\end{figure}

For FF--bubble simulations,
both the FF energy $E$ and off-center distance $\ell$ were discretized,
with $\ell$ being defined as the minimum distance between
the bubble center and the velocity vector of the FF
at the beginning of the simulation.
The energy discretization scheme depended on the specific FF being simulated,
while $\ell$ discretization depended on the bubble radius.
The $\ell$ values were selected to ensure
adequate sampling of the region around the bubble radius $R_b$.
A total of $5,000$ BCA simulations were performed for each configuration.
From these simulations, the number of re-solved Xe atoms was calculated.
This number was then divided by the initial number of Xe atoms in the bubble
in order to compute the re-solved bubble fraction $\chi$.
In other words, if a Xe gas bubble was at the origin $(x, w) = (0, 0)$,
then $\chi(E', \ell')$ would be the re-solved bubble fraction
due to the interaction of the bubble with a FF
originating at $(x, w) = (-D, \ell')$ with an energy $E'$
and moving along the $x$ axis.
The results for bubbles of two different sizes
are shown in Figure \ref{fig:chi}.
The $2$ nm and $64$ nm bubbles are representative
of intragranular and intergranular bubbles, respectively.
The error bars in the figure indicate $2\sigma$ deviations from the mean.

Using the simulation results and interpolation,
it is now possible to estimate any reasonable $\chi(E', \ell')$ value.
If $\mathcal{I}$ is an interpolator that maps $X \rightarrow Y$,
we can use the notation $y' = \mathcal{I}_X (x', X, Y)$ to indicate that
the interpolator returns $y'$ when $x'$ is provided as input.
Any arbitrary $\chi(E', \ell')$ can then be defined as follows:
\begin{align}
	\chi(E', \ell)
	   &= \mathcal{I}_{\mathcal{E}}
	   (E', \mathcal{E}, [\chi(E, \ell)]_{E \in \mathcal{E}}) \\
	\chi(E', \ell')
	   &= \mathcal{I}_{\mathcal{L}}
	   (\ell', \mathcal{L}, [\chi(E', \ell)]_{\ell \in \mathcal{L}})
\end{align}
where $\mathcal{E}$ and $\mathcal{L}$ are the sets of discrete energies
and off-center distances used in the simulations,
and $E$ and $\ell$ are elements of those sets.
In this work,
$\mathcal{I}_{\mathcal{E}}$ is a PCHIP interpolator \cite{fritsch1984},
and $\mathcal{I}_{\mathcal{L}}$ is a linear interpolator.

\subsection{Calculation of \texorpdfstring{$\xi$}{xi}}

\begin{figure}[!ht]
	\centering
	\includegraphics[width=13cm]{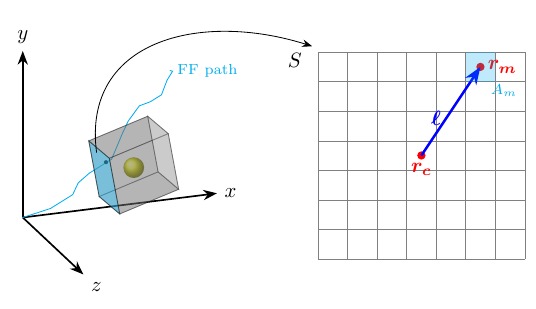}
	\caption{
		\textbf{Illustration of the $\xi$ calculation scheme.}
		$\xi(x, w)$ denotes the re-solved bubble fraction
		for a bubble centered at $(x, w)$ interacting with
		a FF originating at $(0, 0)$ with an initial velocity along the $x$ axis.
		A square surface $S$ with side length $2D$ is constructed
		at a distance $D = R_b + \delta$ distance from the bubble center,
		where the surface normal is oriented at an angle
		$\alpha(x, w)$--the average FF incidence angle--relative to the $x$ axis.
		The surface is discretized into small mesh elements.
		The FF may reach a mesh element $m$
		with a probability per unit area $p_m$ and energy $E$,
		resulting in a re-solved fraction $\chi(E, \ell)$.
		Here, $\ell$ is the distance between
		the element center $r_m$ and the surface center $r_c$.
		The total re-solved fraction $\xi$ is obtained
		by summing the contributions
		from all possible trajectories through the mesh elements.
	}
	\label{fig:surf_mesh}
\end{figure}

Given that $\chi$ values are now known, we now proceed to calculate $\xi$.
Consider a FF at the origin directed along the $x$ axis
and a bubble located at $(x, w)$.
At this position, the FF has an average incidence angle $\alpha(x, w)$.
Since the incidence angle changes minimally with variations in $x$ and $w$,
it is reasonable to assume the incidence angle
in the vicinity of $(x, w)$ is approximately $\alpha(x, w)$.
Next, we identify a point $(x', w')$ such that
it is a distance $D$ away from $(x, w)$
and the line connecting these two points has a slope of $\tan (\alpha(x, w))$.
A surface $S$ perpendicular to this connecting line can then be constructed.
The surface $S$ is square in shape,
with a side length of $2D = 2(R_b + \delta)$.
This surface can then be meshed into small elements,
as illustrated in Figure \ref{fig:surf_mesh}.
The FF has specific probabilities of passing through each mesh element.
If the FF traverses a mesh element $m$
with a probability per unit area $p(r_m)$ and incidence energy $E(r_m)$,
$\xi(x, w)$ can be calculated as:
\begin{align}
	\xi(x, w) &= \sum_{m \in S}
		p(r_m) \frac{A_m}{\cos \alpha(x, w)}
		\chi(E(r_m), ||r_m - r_c||)
	\label{eq:mesh}
\end{align}
where $r_m$ denotes the coordinate of the center of the mesh element,
and $r_c$ denotes the center of the surface $S$.
The probability of the FF passing through the mesh element $m$ is calculated
as the product of $p(r_m)$ and the area of the mesh element $A_m$
projected onto the place perpendicular to the $x$ axis.
It is important to note that all FF trajectories
that do not intersect $S$ are ignored
due to the extremely low probability of the FF or its associated recoils
reaching the bubble.

Although the mesh elements shown in Figure \ref{fig:surf_mesh}
are of constant size,
such a scheme is inefficient for the calculation of $\xi$.
This inefficiency arises because
$R_b$ can be up to $50$ times smaller than $D$.
To address this, we implemented an adaptive meshing scheme.
This scheme ensures
a maximum mesh size of $R_b / 2$ for $\ell \in [0, 2 R_b]$,
and $35$ nm for $\ell \in (2 R_b, D]$.
The upper bounds of the mesh size were selected
so that further refinement would not result in
a relative change in $\xi$ greater than $0.001$.

\begin{figure}[!ht]
	\centering
	\begin{subfigure}{0.49\textwidth}
		\centering
		\caption{}
		\includegraphics
			[width=8cm, trim={0.8cm 0 1.5cm 0.7cm}, clip]
			{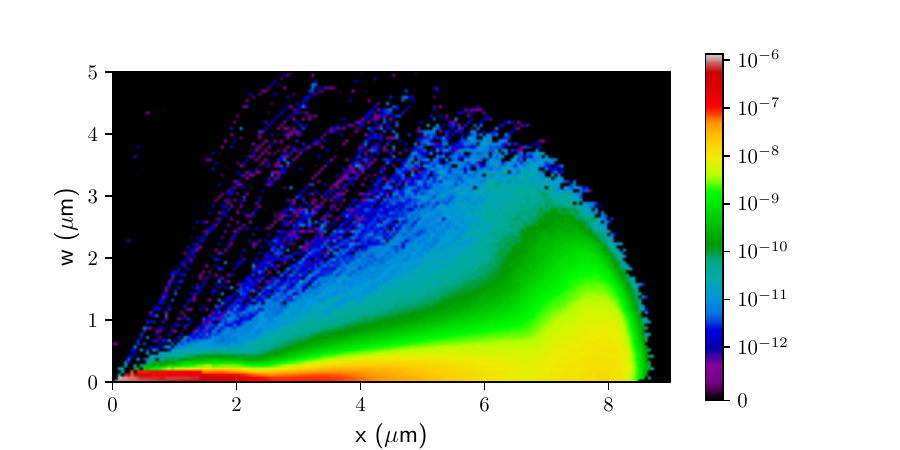}
	\end{subfigure}
	\begin{subfigure}{0.49\textwidth}
		\centering
		\caption{}
		\includegraphics
			[width=8cm, trim={0.8cm 0 1.4cm 0.7cm}, clip]
			{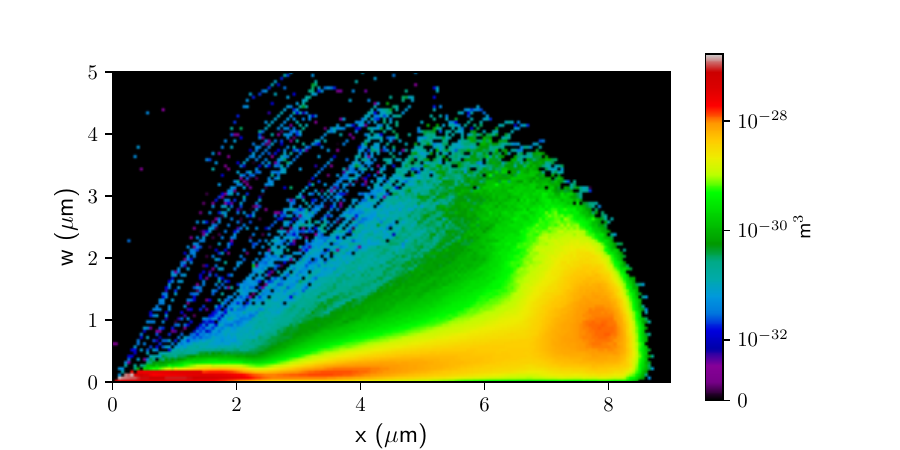}
	\end{subfigure}
	\begin{subfigure}{0.49\textwidth}
		\centering
		\caption{}
		\includegraphics
			[width=8cm, trim={0.8cm 0 1.4cm 0.7cm}, clip]
			{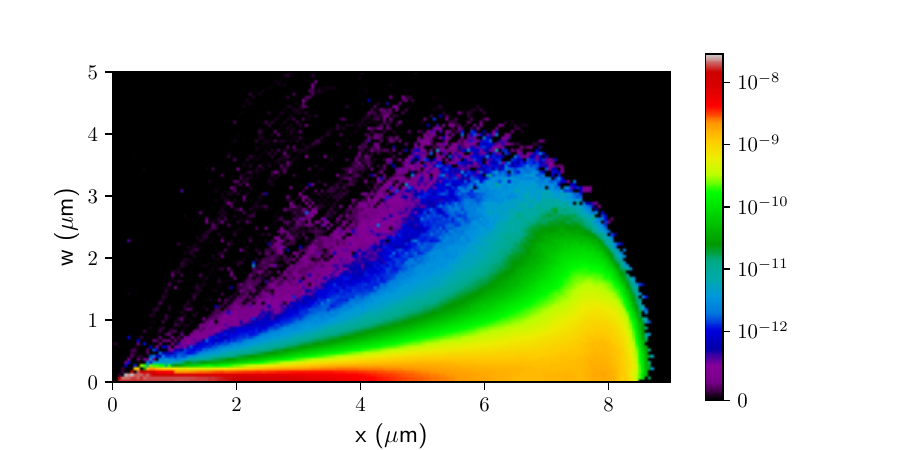}
	\end{subfigure}
	\begin{subfigure}{0.49\textwidth}
		\centering
		\caption{}
		\includegraphics
			[width=8cm, trim={0.8cm 0 1.4cm 0.7cm}, clip]
			{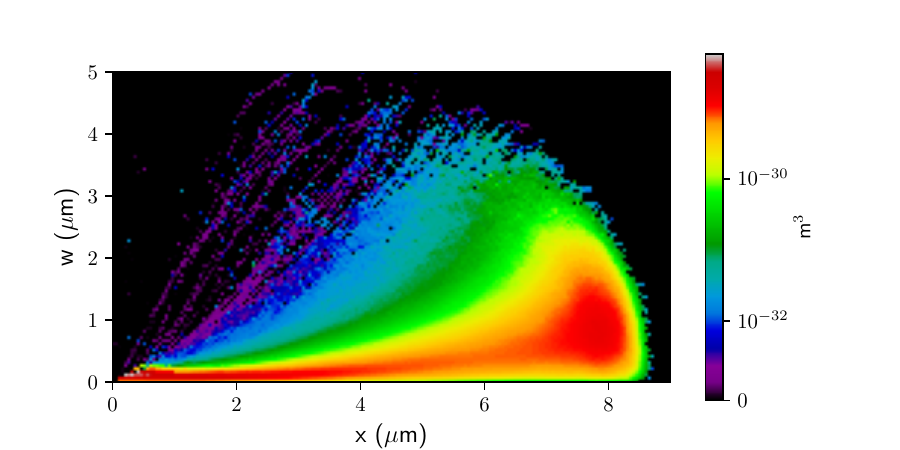}
	\end{subfigure}
	\caption{
		\textbf{Spatial distribution of $\xi$ and $\xi \Delta V$ for \Y FFs.}
		The re-solved bubble fraction $\xi$
		and volume-weighted re-solved bubble fraction $\xi \Delta V$
		are shown for a gas bubble centered at $(0, 0)$
		and a \Y FF at $(x, w)$ pointing toward the $-x$ direction.
		Note that this spatial arrangement is equivalent
		to the one in Figure \ref{fig:coord}
		but represents the inverse of the setup
		shown in Figure \ref{fig:surf_mesh}.
		\textbf{a, b} $\xi$ and $\xi \Delta V$ for a bubble radius of $2$ nm.
		\textbf{c, d} $\xi$ and $\xi \Delta V$ for a bubble radius of $64$ nm.
	}
	\label{fig:xi}
\end{figure}

$\xi$ was calculated for all combinations of FF isotopes and bubble radii.
The $\xi_Y$ values for $2$ nm and $64$ nm bubbles
are shown in Figures \ref{fig:xi}a and \ref{fig:xi}b, respectively.
These $\xi$ profiles can also be interpreted in a way
that simplifies the calculation of the overall re-solution rate $b$.
If a bubble is located at the origin
and a FF positioned at $(x, w)$ is pointing in the $-x$ direction,
the resulting $\xi$ profiles would be exactly the same.
Now, a discretized version of Equation \ref{eq:b} can be formulated as:
\begin{align}
	b / \dot{F}
		&= \sum_{k=Y,I} \sum \xi_k \Delta V \\
		&= \sum_{k=Y,I} \sum_{i,j \in V} \xi_k(x_{i,j}, w_{i,j})
			\left[\pi (w_{i,j+1}^2 - w_{i,j}^2) (x_{i+1,j} - x_{i,j})\right]
	\label{eq:xitob}
\end{align}
where $i, j$ denote the indices of the discrete grid points
where $\xi$ is evaluated.
This is easy to calculate,
because the $\xi$ profiles have already been discretized.
$\xi \Delta V$ profiles are also depicted in Figure \ref{fig:xi}
to show the significant effect of $\Delta V$ on re-solution.

\subsection{Homogeneous re-solution rate}

\begin{figure}[!ht]
	\centering
	\includegraphics[width=8cm]{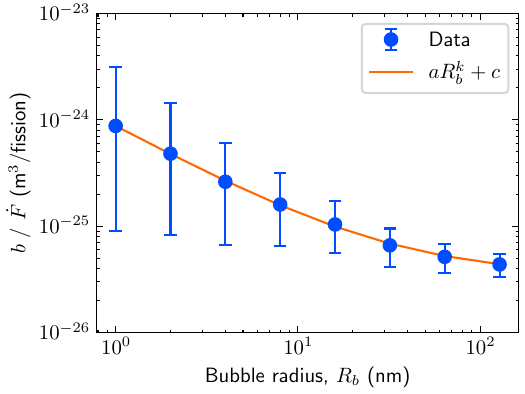}
	\caption{
		\textbf{Homogeneous re-solution rate
		as a function of bubble radius $R_b$ in U-10Mo}.
		The re-solution rate was calculated
		at the equilibrium Xe number density $n_{eq}$
		for bubble radii ranging from $2$ nm to $64$ nm.
		The solid line represents a power-law fit of the form $aR_b^k + c$,
		which describes the data with high precision ($R^2 = 0.99986$).
		Error bars denote $2 \sigma$ deviations from the mean,
		reflecting the uncertainty in $\chi$.
	}
	\label{fig:res}
\end{figure}

The total re-solution rate of a given bubble can be determined by
summing up all the values in $\xi_Y \Delta V$ profiles
due to both \Y and \I.
Figure \ref{fig:res} displays the re-solution rates
for bubbles with radii ranging from $1$ nm to $128$ nm.
The error bars in the figure describe the uncertainty in $b$
that arises solely from the uncertainty in $\chi$.
Since the ion profiles and $\xi$ were computed
using stringent convergence criteria,
it is reasonable to assume that
the majority of the uncertainty originates from $\chi$.
As a result,
the error bars for $b$ approximately correspond to $2\sigma$ deviations.
We also observe vanishingly small deviations for larger bubbles,
simply because the probability of an interaction
between a FF and larger bubbles is higher.

While linear interpolations are sufficient
to estimate re-solution rates for arbitrary bubble radii,
an approximate analytical function may be more suitable
for use in higher-length-scale models.
To this end, we propose the following functional form:
\begin{align}
	b / \dot{F} = a R_b^k + c
\end{align}
where $a = \num{8.43e-25}, k = -0.926$ and $c = \num{3.46e-26}$.
$R_b$ is in nm, and $b / \dot{F}$ is in m$^3$/fission.
The functional fit has
a root mean squared error of $\num{3.17e-27}$ m$^3$/fission
and a $R^2$ score of $0.99986$.

\subsection{Effect of bubble pressure}

\begin{figure}[!ht]
	\centering
	\includegraphics[width=8cm]{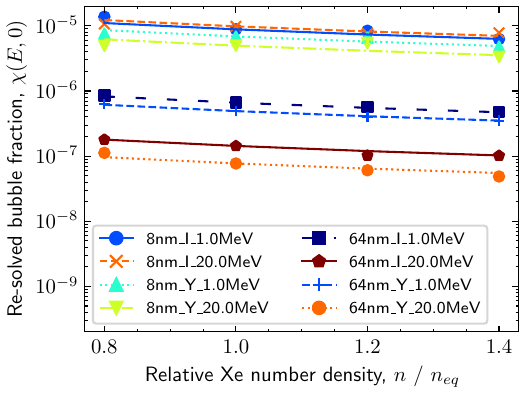}
	\caption{
		\textbf{Effect of Xe number density ($n$)
		on re-solved bubble fraction ($\chi$).}
		BCA simulations of FF interactions with under- and over-pressurized bubbles
		reveal an inverse relationship between $\chi$ and $n$.
		The lines represent the fit $\chi / \chi_{eq} = n_{eq} / n$,
		which accurately describes the data ($R^2 = 0.97$).
		This relationship arises because
		the number of re-solved Xe atoms remains nearly invariant
		regardless of the bubble's Xe number density.
		Data were obtained from $5,000$ independent BCA simulations
		per configuration.
	}
	\label{fig:pres}
\end{figure}

The effect of pressure on re-solution was investigated
by varying the Xe number density within bubbles of radii $8$ nm and $64$ nm.
FF--bubble interaction simulations were conducted
for \Y and \I ions with energies of $1$ MeV and $20$ MeV.
The limited scope of simulations investigating the effect of bubble pressure
is intended to show general trends in the re-solution behavior
of under- and over-pressurized bubbles.
The resulting $\chi$ values that were calculated from these simulations
are plotted in Figure \ref{fig:pres},
showing an inverse relationship between $\chi$ and $n$.
The equation $\chi / \chi_{eq} = n_{eq} / n$,
which was used to make the lines in Figure \ref{fig:pres},
fits the data with an $R^2$ score of $0.97$.
To understand the origin of this inverse relationship,
further analysis of the simulation data was performed.
It was found that the number of re-solved Xe atoms is almost an invariant
with respect to the initial Xe number density in the bubble.
Since $\chi$ is defined as the ratio of the number of re-solved Xe atoms
to the total number of Xe atoms in the bubble,
an inverse relation between $\chi$ and $n$ naturally arises.
One possible explanation for the invariance of the number of re-solved atoms
is that the re-solution behavior is primarily affected by
FF--bubble interactions close to the bubble surface,
and the gas bubble surface area is independent of $n$.

Up to this point, the symbol $b$ has been used to denote the re-solution rate
at the equilibrium Xe number density $n_{eq}$.
Moving forward, $b_{eq}$ will specifically refer to the re-solution rate
at equilibrium Xe density,
while $b$ will be reserved for the general re-solution rate.
The re-solved bubble fraction $\chi$ is related to $\xi$
through Equation \ref{eq:mesh},
and $\xi$ is related to $b$ through Equation \ref{eq:xitob}.
These relations can thus be used
to relate the re-solution rate directly to the Xe number density:
\begin{align}
	b &\propto \xi \propto \chi \\
	b / b_{eq} &= n_{eq} / n \\
	b &= \left( a R_b^k + c \right) \left( \frac{n_{eq}}{n} \right) \dot{F}
	\label{eq:model}
\end{align}
where $R_b$ is in nm, $n$ is in m$^{-3}$,
$\dot{F}$ is in fission m$^{-3}$ s$^{-1}$, and $b$ is in s$^{-1}$.
Equation \ref{eq:model} now describes the fission gas bubble re-solution rate
as a function of bubble size, bubble pressure, and fission rate.

\section{Discussion}

\begin{figure}[!ht]
	\centering
	\includegraphics[width=8cm]{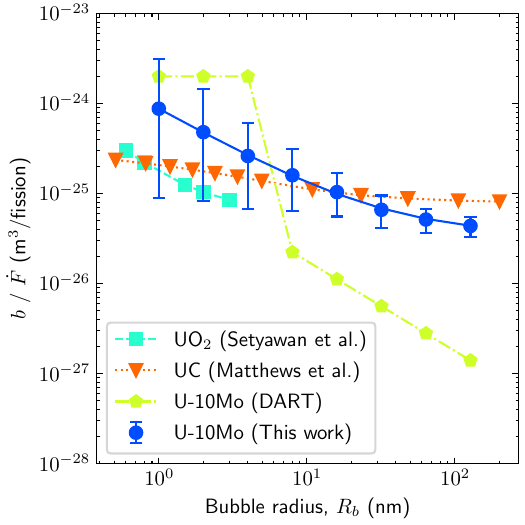}
	\caption{
		\textbf{Comparison of the re-solution rates in UO$_2$, UC, and U-10Mo.}
		The re-solution rates in various nuclear materials are shown
		as a function of bubble radius $R_b$.
		For UO$_2$, the re-solution rate is dominated by the heterogeneous mechanism,
		which is driven by the thermal spikes
		resulting from the electronic stopping of FFs.
		In contrast, for UC and U-10Mo, re-solution is primarily governed by
		the homogeneous mechanism (nuclear stopping)
		since their higher electronic conductivities suppress thermal spike formation.
		The piecewise re-solution model for U-10Mo employed in the DART code
		is also included for comparison.
	}
	\label{fig:comp}
\end{figure}

Figure \ref{fig:comp} compares the re-solution rate calculated in this work
against the literature values for UO$_2$, UC, and U-10Mo.
Setyawan et al. performed MD simulations of gas bubbles in UO$_2$,
with the radii ranging from $0.6$ nm to $3$ nm \cite{setyawan2018}.
The slope of $\ln b$ vs. $\ln R_b$ in UO$_2$
is similar to that observed in our work,
albeit the re-solution rate in UO$_2$ is nearly
one order of magnitude lower than in U-10Mo.
Matthews et al. conducted BCA simulations to evaluate the re-solution rate
for a wide range of bubble radii in UC \cite{matthews2015}.
The slope of $\ln b$ vs. $\ln R_b$ in UC is flatter than in U-10Mo.
Interestingly, an intersection between the two rates is also observed:
the re-solution rate is higher in U-10Mo for smaller bubbles,
but higher in UC for larger bubbles.
Despite these differences,
the re-solution rates in UO$_2$, UC, and U-10Mo (this work)
are remarkably close,
with the variation often being less than one order of magnitude.
We could not compare our results against U-Zr
because Mao et al. simulated all interactions at a fixed distance of $2$ $\mu$m
between the FFs and the bubbles \cite{mao2025}.
Thus, their data do not directly lead to an overall re-solution rate.

Finally, we compare the computed re-solution rate from our work
against the existing re-solution model of U-10Mo as implemented in DART.
The model is defined as follows:
\begin{align}
	b_{dart} &= b_0 \cdot \dot{F} \cdot G \\
	G &=
	\begin{cases}
		1 & ,R_b \leq \lambda \\
		1 - (\frac{R_b - R_{resol}}{R_b})^3
		  & ,R_b > \lambda
	\end{cases}
\end{align}
where $b_0$ is the bubble destruction probability,
and $G$ is a piecewise function representing
different re-solution modes for small and large gas bubbles.
In the piecewise function $G$,
$\lambda$ is the gas-atom knock-out distance
and $R_{resol}$ is the thickness of the annular region
within which all gas-atoms are considered to be knocked out.
The parameters $b_0$, $\lambda$, and $R_{resol}$ are treated as adjustable,
and were optimized to the following values:
$b_0 = \num{2e-18}$ cm$^3$, $\lambda = \num{5e-7}$ cm,
and $R_{resol} = \num{3e-9}$ cm.
It is important to note that
since the parameters $\lambda$ and $R_{resol}$ are not coupled,
the re-solution rate exhibits a discontinuity
when the bubble radius is $\lambda$.
According to Ye et al. \cite{ye2023},
this is due to the strong trapping effects of grain boundaries
on intergranular bubbles.

The re-solution rate calculated in this work is slightly lower than
the DART model prediction for intragranular bubbles
but significantly higher for intergranular bubbles.
In fact, the difference can be as large as two orders of magnitude
for intergranular bubbles.
This large discrepancy potentially arises from the mixing of trapping effects
in the re-solution rate in the DART model.
However, the trapping effect of the grain boundaries
is negligible in the time frame of collision cascades.
It is therefore recommended that
the re-solution rate and trapping rate be implemented as separate processes
in the higher-length-scale models of gas bubble evolution.
This separation would provide greater flexibility
for modeling the complex behavior of gas bubbles in the fuel
under different contexts.
For instance, the grain boundary diffusion coefficient of Xe in U-10Mo can be
$15$ orders of magnitude higher than the intrinsic diffusion coefficient
at approximately 600 K \cite{hasan2024gb}.
In a gas bubble evolution model,
the Xe trapping rate for intergranular bubbles could therefore be set
$15$ times higher than for intragranular bubbles,
while the re-solution rate could be independently specified
using the analytical model developed in this work.
By decoupling these two rates,
the re-solution rate would not be confounded with the trapping rate,
enabling more accurate predictions of gas bubble behavior in U-10Mo.

\section{Conclusions}

This study combined BCA and MD simulations to provide
a comprehensive, physics-based understanding
of the Xe gas bubble re-solution rate in U-10Mo nuclear fuel.
Our findings reveal that homogeneous re-solution driven by nuclear stopping
is the only active mechanism in U-10Mo,
as heterogeneous re-solution via electron stopping is highly improbable.
A systematic approach was employed,
involving profiling FF behavior,
evaluating the interactions between FFs and Xe gas bubbles,
and integrating the results into a physical model.
The computed re-solution rate spanned
from $\num{4.4e-26} \dot{F}$ s$^{-1}$ for the largest intergranular bubble
to $\num{8.8e-25} \dot{F}$ s$^{-1}$ for the smallest intragranular bubble,
with $\dot{F}$ expressed in fission/m$^3$/s.
Furthermore, our simulations demonstrated an inverse relationship
between the re-solution rate and Xe number density,
indicating that higher bubble pressures suppress re-solution.
These results provide critical insights into
the fundamental mechanisms underpinning gas bubble behavior in nuclear fuels,
and establish a robust foundation for higher-length-scale models of U-10Mo.

\section{Data availability}

All data described in this work can be found
at \url{https://osf.io/eubd6/}.

\section{Code availability}

The RustBCA fork containing
the implementation of \texttt{SPHEREINCUBOID} geometry
is available at \url{https://github.com/ATM-Jahid/RustBCA}.

\section{Acknowledgements}

This work was supported by the U.S. Department of Energy,
Ofﬁce of Material Management and Minimization,
National Nuclear Security Administration,
under DOE-NE Idaho Operations Ofﬁce Contract DE-AC07-05ID14517.
This manuscript has been authored in part
by Battelle Energy Alliance, LLC with the U.S. Department of Energy.
The publisher, by accepting the article for publication,
acknowledges that the U.S. Government retains
a nonexclusive, paid-up, irrevocable, worldwide license
to publish or reproduce the published form of this manuscript,
or allow others to do so, for U.S. Government purposes.
This research made use of the resources of
the High Performance Computing Center at Idaho National Laboratory,
which is supported by the Ofﬁce of Nuclear Energy
of the U.S. Department of Energy and the Nuclear Science User Facilities.

\section{Author contributions}

A.H. conceptualized the study,
performed the simulations, and analyzed the data.
A.H. also prepared the original draft of the manuscript.
L.M. and M.S. contributed to the MD simulation design
and assisted with the initial draft.
B.B. provided supervision, project administration,
and funding acquisition.
B.B. also performed the final review and editing of the manuscript.

\section{Competing interests}

The authors declare no competing interests.

\bibliographystyle{unsrt}
\bibliography{ref.bib}

\end{document}